\documentclass[envcountsame]{llncs}
    \usepackage{amsmath}
    \usepackage{amsxtra}
    \usepackage{amstext}
    \usepackage{amssymb}
    \usepackage{latexsym}
    \usepackage{dsfont} 
\newcommand{\GF}[1]{{\mathbb F}_{#1}}

\begin{document}
\title{Solving 
$X^{2^{3n}+2^{2n}+2^{n}-1}+(X+1)^{2^{3n}+2^{2n}+2^{n}-1}=b$ in
$\GF{2^{4n}}$}
\author{Kwang Ho Kim\inst{1,2} \and Sihem Mesnager\inst{3}}

\institute{Institute of Mathematics, State Academy of Sciences,
Pyongyang, Democratic People's Republic of Korea\\
\email{khk.cryptech@gmail.com} \and PGItech Corp., Pyongyang, Democratic People's Republic of Korea\\ \and Department of Mathematics, University of Paris VIII, F-93526 Saint-Denis, University Sorbonne Paris Cit\'e, LAGA, UMR 7539, CNRS, 93430 Villetaneuse and Telecom Paris, Polytechnic Institute of Paris, 91120 Palaiseau, France.\\
\email{smesnager@univ-paris8.fr}\\} \maketitle

\begin{abstract}
This article determines all the solutions in the finite field $\GF{2^{4n}}$ of the equation
$x^{2^{3n}+2^{2n}+2^{n}-1}+(x+1)^{2^{3n}+2^{2n}+2^{n}-1}=b$. Specifically, we explicitly determine the set of $b$'s for which the equation has $i$ solutions for any positive integer $i$. Such sets, which depend on the number of solutions $i$, are given explicitly and expressed nicely, employing the absolute trace function over $\GF{2^{n}}$, the norm function over $\GF{2^{4n}}$ relatively to $\GF{2^{n}}$ and the set of $2^n+1$st roots of unity in $\GF{2^{4n}}$. The equation considered in this paper comes from an article by Budaghyan et al. The authors have been interested in investigating approaches for obtaining alternative representations for functions from the known infinite APN families. In particular, they have been interested in determining the differential spectrum of some power functions among them is the one $F(x)= x^{2^{3n}+2^{2n}+2^{n}-1}$ defined over $\GF{2^{4n}}$. The problem of the determination of such spectrum has led to a conjecture (Conjecture 27 in the preprint (2020) \cite{BCCDK20} for which an updated version will appear in 2022 at the IEEE Transactions Information Theory) stated by Budaghyan et al.  

As an immediate consequence of our results, we prove that the above equation has $2^{2n}$ solutions for one value of $b$, $2^{2n}-2^n$ solutions for $2^n$ values
of $b$ in $\GF{2^{4n}}$ and has at most two solutions for all remaining points $b$, leading to complete proof of the conjecture raised by Budaghyan et al. We highlight that the recent work of Li et al., in \cite{Li-et-al-2020} gives the complete differential spectrum of $F$ and also gives an affirmative answer to the conjecture of Budaghyan et al. However, we emphasize that our approach is interesting and promising by being different from Li et al. Indeed, on the opposite to their article, our technique allows determine ultimately the set of $b$'s for which the considered equation has solutions as well as the solutions of the equation for any $b$ in $\GF{2^{4n}}$. 

\end{abstract}


\section{Introduction}

Let $n$ be a positive integer. Let $\GF{Q}$ be the finite field with
$Q$ elements. Let $q=2^n$ and $d=q^3+q^2+q-1$. In the article, \cite{BCCDK20} (for which an updated version will appear in 2022 at the IEEE Transactions Information Theory),  Budaghyan, Calderini, Carlet, Davidova, and Kaleyski have presented a promising approach based on considering alternative representations of infinite almost perfect nonlinear (APN) monomials that could lead to the resolution of fundamental problems related the known families of power APN functions. Those functions are very important in symmetric cryptography since they contribute to an optimal resistance against differential cryptanalysis, a powerful attack employed against block ciphers. A lot of attention and efforts have been made, as can be seen notably in the nice and complete Chapter 11 in the recent book \cite{Carlet-book-2021}. The following conjecture is among those fundamental problems presented very recently by Budaghyan et al. It concerns the  number of solutions to the equation

\begin{equation}
  \label{ori_eq}
  X^d + (X+1)^d = b
\end{equation}
in $\GF{q^4}$  (where  $b\in\GF{q^4}$).

\begin{conjecture}[Conjecture 27 of \cite{BCCDK20}]\label{conjecture-BCCDK20}
  Equation~(\ref{ori_eq}) has $q^2$ solutions for one value of
  $b$, it has $q^2-q$ solutions for $2^n$ values of $b$ and has at
  most $2$ solutions for all remaining points $b$.
\end{conjecture}

The above conjecture is based on computational results reported by  Budaghyan et al. in \cite{BCCDK20} and its original interest  comes from the determination of the differential spectrum of the power function $x^d$ where  $d=2^{3n}+2^{2n}+2^{n}-1$ over $\GF{2^n}$ (see more details in Section \ref{Motivation}).
This  article is completely devoted to proving this conjecture and even
more since we show a more precise statement (Theorem \ref{thm:main-result}).
More specifically,  we completely determine for which $b\in\GF{q^4}$
Equation~(\ref{ori_eq}) has no solution, one solution, $q^2-q$
solutions or $q^2$ solutions. 
For our approach, we  shall denote by $\mu_m$ the set of  (non-zero) elements of $\GF{q^4}$ which are $m$-st roots of unity, that is
 $\mu_m:=\{x\in \GF{q^4}\mid x^m=1\}$. 
 The following theorem is the main result of the paper. 
\begin{theorem}\label{thm:main-result}
  Let $b\in\GF{q^4}$. Let $d=q^3+q^2+q-1$. We define $\mathfrak{S}_2$ as follows
  $$\mathfrak{S}_2:=\left\{b\in \GF{q^4}\setminus \GF{q^2} \mid
    \mathbf{Tr}_1^{n}\left(\mathbf{N}_n^{4n}\left(\frac{b^{q+1}+1}{b+b^{q^2}}\right)\right)=1\right\},$$ 
    where  $\mathbf{Tr}_1^{n}$ denotes the (absolute) trace function over $\GF{2^n}$ and  $\mathbf{N}_n^{4n}$ denotes the norm function of the extension fields $\GF{2^{4n}}/\GF{2^{n}}$.
    
    Then,
 the number of solutions to the equation $x^d+(x+1)^d=b$ in $\GF{q^4}$ is
  equal to
  \begin{enumerate}
  \item\label{item:1} $q^2$ if $b=1$,
  \item\label{item:2} $q^2-q$ if $b\in\mu_{q+1}$,
  \item\label{item:3} $2$ if $b\in\mathfrak{S}_2\setminus\GF{q^2}$.
  \end{enumerate}
   Otherwise, for the other values of $b$, the equation $x^d+(x+1)^d$
 has no solution in $\GF{q^4}$.
 \end{theorem}

Note that Li, Wu, Zeng, and Tang have been interested in the above conjecture and proved it in \cite{Li-et-al-2020}. We emphasize our algebraic approach, which leads, in particular, to solve this conjecture is different from the one presented by Li et al. In addition, we go beyond since we also determine ultimately the set of $b$'s (involving $\mathfrak{S}_2$ and $\mu_{2^n+1}$) for which the considered equation has solutions as well as the solutions of the equation for any $b$ in $\GF{2^{4n}}$.

The proof of each item is given in Section \ref{main-result} with separate subsections (Subsections
\ref{sec:proof-item-refitem:1} to \ref{sec:proof-item-refitem:3}).
Furthermore, besides proving Conjecture 27 in
\cite{BCCDK20}, our approach  allows to solve
Equation~(\ref{ori_eq}) for any
$b$ (see Lemma~\ref{Fcase} for the first item,
Remark~\ref{rem-Scase} for the second item and Remark~\ref{rem:Tcase}
for the third item).
$q^2-q$ solutions (see Lemma~\ref{Fcase} for the first item,
Remark~\ref{rem-Scase} for the second item and Remark~\ref{rem:Tcase}
for the third item).
The rest of this paper is organized as follows. In Section
\ref{sec-prel}, we first fix our notation and present some
basic notions and a few known results helpful in the technical part of the paper.
Next, in Section \ref{Motivation} we explain a connection of this contribution to a problem related to symmetric cryptography that motivated the Budaghyan et al. (\cite{BCCDK20}) by stating the considered conjecture. Section
\ref{main-result} is devoted to the proof of our main result.
Finally, Section \ref{Sec-Conclusion} concludes the paper.

\section{Preliminaries}\label{sec-prel}

\subsection{Notation and basic notions related to  finite fields}

The cardinality of a finite set $A$ is denoted by $\#A$.
Let $n$ be a positive integer and $Q$ be a prime power of $2$. Let $\mathbb F_Q$  be a prime field of characteristic $2$  of cardinality $Q$.
 We denote by $\mathbb F_{Q}^*$ the multiplicative cyclic group of non-zero elements of the finite field $\mathbb F_{Q}$. 
 If $Q=2^n$ then  $\mathbb F_Q$ is an extension field of degree $n$ over $\mathbb F_2$ and also a vector space of dimension $n$ over $\mathbb F_2$. 
Given $x\in\GF{Q}$ we shall denote $\sqrt x=x^{\frac{1}{2}}$ the image of $x$ by the inverse of the Frobenius  map over $\GF{Q}$. Note that we will use the two symbols `` $\sqrt \cdot$ "and ``$\cdot^{\frac{1}{2}}$ " to make certain expressions less cumbersome.
 Let $k$ and $l$ be two integers such that $l$ is a divisor of $k$. Define the norm $\mathbf{N}_l^k$
and trace $\mathbf{Tr}_l^k$ mappings from $\GF{2^k}$ to $\GF{2^l}$ by 
$$\mathbf{N}_l^k(x):=\prod_{i=0}^{\frac kl-1}x^{2^{li}},\quad \mathbf{Tr}_l^k(x):=\sum_{i=0}^{\frac kl-1}x^{2^{li}},$$
respectively.




 
 \subsection{Some elementary results}
We first recall a classical  result related  to a composition of nonzero elements of $\GF{2^m}$.
\begin{proposition}[{\cite[Proposition 1]{KM19}}]\label{Aprop}
  \label{prop:decomposition}
  Let $m$ be a positive integer. Then, every element $z$ of
  $\GF{2^m}^*:=\GF{2^m}\setminus\{0\}$ can be written twice as
  $z=c+\frac 1c$ where $c\in\GF{2^m}^\star:=\GF{2^m}\setminus \GF{}$
  if $\mathbf{Tr}_1^m(\frac{1}{z})=0$ and
  $c\in \mu_{2^m+1}^{\star}:=\{\zeta\in\GF{2^m}\mid
  \zeta^{2^m+1}=1\}\setminus \{1\}$ if
  $\mathbf{Tr}_1^m(\frac{1}{z})=1$.
\end{proposition}
Next, we present below a decomposition of $\GF{q^4}$ that comes
from  that the integers $q-1$, $q+1$ and $q^2+1$ are pairwise coprime
and that 
$$\gcd(q^3+q^2+q-1, q^4-1)=\gcd(q^3+q^2+q-1, (q-1)(q+1)(q^2+1))=1.$$
\begin{lemma}
 $\GF{q^4}^*$ can be composed  as $\GF{q^4}^*=\mu_{q-1}\cdot\mu_{q+1}\cdot \mu_{q^2+1}$.
\end{lemma}

\section{A motivation in symmetric cryptography}\label{Motivation} 
In this section we explain one motivation in solving the equation
$x^{2^{3n}+2^{2n}+2^{n}-1}+(x+1)^{2^{3n}+2^{2n}+2^{n}-1}=b$ in
$\GF{2^{4n}}$. The equation which will be considered in the paper has an interest in the context of differential uniformity analysis of vectorial Boolean functions (also called S-boxes in cryptography) from $\mathbb{F}_{2^n}$ to $\mathbb{F}_{2^n}$ viewed as important components in symmetric cryptosystem because of their relevance in the construction of S-boxes in block ciphers. In fact, for resistance of cryptosystem against differential attacks (\cite{Biham-Shamir-1991}), S-boxes should have low differential uniformity (\cite{SM-NK}). The notion of the differential uniformity of a function $F$ was introduced by Nyberg (\cite{Nyberg-93,SM-NK}) as an important parameter that measures the resistance of the block cipher involving $F$  against differential cryptanalysis. Precisely, $F$ has differential uniformity $d$ if $d= max_{a \in\mathbb{F}_{2^n}^\star, b\in \mathbb{F}_{2^n}}\#\{x\in \mathbb{F}_{2^n} \mid F(x)+F(x+a)=b\}.$ A function with differential uniformity $2$ is called almost perfect nonlinear (APN) on $\mathbb{F}_{2^n}$. The differential uniformity is always a multiple of $2$, so APN functions have the lowest possible differential uniformity and give the best protection against differential attacks. As a special class of functions over finite fields, power functions, namely, monomial functions, have been extensively studied in the last decades due to their simple algebraic form and lower implementation cost in a hardware environment. Very recently, Budaghyan et al. have investigated in \cite{BCCDK20} a novel approach for handling some APN efficiently through an interesting analysis related to their representations and presented some computational data on the differential spectra of power functions $F(x)=x^d$ with $d=\sum_{i=1}^{k-1}2^{in}-1$ where $n$ and $k$ are two positive integers. Power functions  have been extensively studied in the last decades due to their simple algebraic form and lower implementation cost in a hardware environment. Also, it is worth noting that this class of power functions includes some famous functions as special cases. For $n=1$, then $F(x)=x^d$ coincides with the well-known inverse function, which is either APN or $4$-differential uniform and has been widely used in practical cryptosystems. If $k=2$ then $F(x)=x^d$
If $k=3$, then $F(x)=x^d$ is the well-known Kasami (\cite{CKM2021,D99,Kasami}). When $k=5$, then $F(x)=x^d$ is the well-known Dobbertin (\cite{D99-1}). The two former power functions are important APN functions among the six infinite families of power APN functions are considered the oldest known instances of APN functions, and it was conjectured in 2000 that they exhaust all possible power APN functions. The differential spectrum of $F(x)=x^d$ when $k=4$ was determined by   Li, Wu, Zeng and Tang (\cite{Li-et-al-2020}) in 2020. Their main result was given by Theorem 1 in (\cite{Li-et-al-2020}) and motivated by the conjecture (Conjecture 27) by Budaghyan et al. in \cite{BCCDK20} directly related to the differential uniformity of the power function $x^d$ where $d=2^{3n}+2^{2n}+2^{n}-1$. Li et al. handed this conjecture and confirmed its validity. Their interesting result was sufficient to determine the differential spectrum of the power function $F$. In this paper, we propose an approach to solve the conjecture of Budaghyan et al. mentioned above but also go beyond by determining ultimately the set of $b$'s for which the considered equation has solutions as well as the solutions of the equation for any $b$ in $\GF{2^{4n}}$. 



\section{Proof of Theorem \ref{thm:main-result}}\label{main-result}

We shall adopt the following organisation of the paper. The whole goal of the present section is to prove Theorem \ref{thm:main-result}. We attributed each subsection to the proof of each assertion given in Theorem \ref{thm:main-result}.

\subsection{Proof of Item~\ref{item:1}}
\label{sec:proof-item-refitem:1}

We prove the following.
\begin{lemma}\label{Fcase}
We have
\[
 \{x\in \GF{q^4}\mid
x^{q^3+q^2+q-1}+(x+1)^{q^3+q^2+q-1}=1\}=\GF{q^2}.
\]
\end{lemma}
\begin{proof}
  Every element $x$ of $\GF{q^2}$ satisfy
  $x^{q^3+q^2+q-1}+(x+1)^{q^3+q^2+q-1}=1$ because
  $x^{q^3+q^2+q-1}=x^{q+1+q-1}=x^{2q}$ and
  $x^{q^3+q^2+q-1}+(x+1)^{q^3+q^2+q-1}=x^{2q}+(x+1)^{2q}=1$.
  
  Conversely, let $x\in\GF{q^4}$ such that
  $x^{q^3+q^2+q-1}+(x+1)^{q^3+q^2+q-1}=b$ i.e.
  $\frac{\mathbf{N}_n^{4n}(x)}{x^2}+\frac{\mathbf{N}_n^{4n}(x+1)}{(x+1)^2}=1$,
  that is,
  $$x^2+ux+v=0$$ where
  $u=\left(1+\mathbf{N}_n^{4n}(x)+\mathbf{N}_n^{4n}(x+1)\right)^{\frac{1}{2}}$
  and $v=\mathbf{N}_n^{4n}(x)^{\frac{1}{2}}$ are both in $\GF{q}$. The
  above equation being a quadratic equation over $\GF{q}$, that
  implies that $x\in \GF{q^2}$.\qed
\end{proof}

\subsection{Proof of
  Item~\ref{item:2}}\label{sec:proof-item-refitem:2}

We shall need the following Lemma in the proof of Item~\ref{item:2}. 

\begin{lemma}
\[
(1+\mu_{(q-1)(q^2+1)})\cap \mu_{(q-1)(q^2+1)}=\GF{q}\setminus
\GF{2}.
\]
\end{lemma}
\begin{proof}
  Firstly observe that
  $\GF{q}\setminus \GF{2}\subset (1+\mu_{(q-1)(q^2+1)})\cap
  \mu_{(q-1)(q^2+1)}$ because
  $\GF{q}\setminus \{0\}\subset \mu_{(q-1)(q^2+1)}$.

  Now let us prove the converse inclusion. Let
  $y,z\in \mu_{(q-1)(q^2+1)}$ satisfying
\begin{equation}\label{1eq}
 1+y=z.
\end{equation}
Raising \eqref{1eq} to the $q-$th power and adding the so-obtained
equation to~(\ref{1eq}) gives
\begin{equation}\label{20220}
y(1+y^{q-1})=z(1+z^{q-1}).
\end{equation}
Again raising \eqref{20220} to $q^2-$th power yields
$y^{q^2}(1+y^{q^2(q-1)})=z^{q^2}(1+z^{q^2(q-1)}).$ Since
$y^{(q-1)(q^2+1)}=z^{(q-1)(q^2+1)}=1$, it follows
$y^{q^2}(1+y^{-(q-1)})=z^{q^2}(1+z^{-(q-1)})$, i.e.
\begin{equation}\label{20221}
y^{q^2-q+1}(1+y^{q-1})=z^{q^2-q+1}(1+z^{q-1}).
\end{equation}
Let us assume that $y\notin \GF{q}$ and so $y(1+y^{q-1})\neq0$. Then,
by dividing \eqref{20221} by \eqref{20220} side by side we get
$y^{q(q-1)}=z^{q(q-1)}$, i.e. $y^{q-1}=z^{q-1}$. Then \eqref{20220}
gives $y=z$ which is a contradiction to \eqref{1eq}. Consequently,
$y\in \GF{q}$ and $z\in \GF{q}$.\qed
\end{proof}

Thanks to the above lemma, we prove.

\begin{lemma}\label{Scase} Let $b\in \GF{q^2}\setminus \{1\}$.
\[
\#\{x\in \GF{q^4}\mid
x^{q^3+q^2+q-1}+(x+1)^{q^3+q^2+q-1}=b\}=\begin{cases}q^2-q, \text{
if $b\in \mu_{q+1}$}\\
0, \text{ otherwise.}
\end{cases}
\]
\end{lemma}
\begin{proof} Let $x\in \GF{q^4}$ and $b\in \GF{q^2}\setminus
\{1\}$. Then, by raising to the $q^2-$th power the both sides of
\eqref{ori_eq}, we get
\begin{equation}\label{q_powering_eq}
x^{q^3-q^2+q+1}+(x+1)^{q^3-q^2+q+1}=b.
\end{equation}
Adding the equalities \eqref{ori_eq} and \eqref{q_powering_eq} gives
$$
x^{q^3+q^2+q-1}+x^{q^3-q^2+q+1}+(x+1)^{q^3+q^2+q-1}+(x+1)^{q^3-q^2+q+1}=0,
$$
i.e.
$$
\frac{\mathbf{N}_n^{4n}(x)(x+x^{q^2})^2}{x^{2q^2+2}}+\frac{\mathbf{N}_n^{4n}(x+1)(x+x^{q^2})^2}{(x+1)^{2q^2+2}}=0.
$$
By Lemma~\ref{Fcase}, when $b\neq 1$, any $x\in \GF{q^4}$ satisfying
\eqref{ori_eq} is not in $\GF{q^2}$, i.e. $x+x^{q^2}\neq 0$. Hence we
get
$\frac{\mathbf{N}_n^{4n}(x)}{x^{2q^2+2}}+\frac{\mathbf{N}_n^{4n}(x+1)}{(x+1)^{2q^2+2}}=0$,
or
$\left(1+\frac{1}{x}\right)^{2q^2+2}=\mathbf{N}_n^{4n}\left(1+\frac{1}{x}\right)$,
which can be rewritten as
$\left(1+\frac{1}{x}\right)^{(q-1)(q^2+1)}=1.$ Let $1+\frac{1}{x}=zt$
for some $(z,t)\in \mu_{q-1}\times\mu_{q^2+1}$.  Substituting 
\begin{equation}\label{x0}
x=\frac{1}{1+zt}
\end{equation} to \eqref{ori_eq} yields
$$\left(\frac{1}{1+zt}\right)^{q^3+q^2+q-1}\left(1+(zt)^{q^3+q^2+q-1}\right)=b.$$
Since
$1+(zt)^{q^3+q^2+q-1}=1+z^2t^{-2}=1+(zt)^{2q^2}=(1+zt)^{2q^2}$, we rewrite the above
equality as
\begin{equation}\label{Cond}
(1+zt)^{(q-1)(q^2+1)}=b^{-1}.
\end{equation} 
It follows that, Equation \eqref{ori_eq} has a solution in
$\GF{q^4}$ for $b\in \GF{q^2}\setminus \{1\}$, $b\in \mu_{q+1}$.

Conversely, let us assume $b\in \mu_{q+1}$. We have to count the
number of pairs $(z,t)\in \mu_{q-1}\times\mu_{q^2+1}$ satisfying
\eqref{Cond}.  For that, write
\begin{align*}
&(1+zt)^{(q-1)(q^2+1)}=\left((1+zt)^{q^2+1}\right)^{q-1}=\left((1+zt)(1+zt^{-1})\right)^{q-1}
\\&=\left(1+z^2+z(t+t^{-1})\right)^{q-1}=\left(z+z^{-1}+t+t^{-1}\right)^{q-1}.
\end{align*}
Observe that the $(q-1)-$th powering acts on $\mu_{q+1}$ as the
$(-2)-$th powering, and therefore, letting $c=b^{\frac{1}{2}}$, one
has
\begin{equation}\label{eq3}
t+t^{-1}=z+z^{-1}+c\cdot w, w\in \GF{q}^*.
\end{equation}
 Let $T=\frac{1}{t+t^{-1}}$.  By Proposition~\ref{Aprop},
$t\in \mu_{q^2+1}$ if and only if
\begin{equation}\label{maincondition}
\mathbf{Tr}_1^{2n}\left(\frac{1}{z+z^{-1}+c w}\right)=1.
\end{equation} Let $z+z^{-1}=0$, i.e. $z=1$. Then, since $
\mathbf{Tr}_1^{2n}\left(\frac{1}{cw}\right)=\mathbf{Tr}_1^{n}\left(\mathbf{Tr}_n^{2n}\left(\frac{1}{cw}\right)\right)
=\mathbf{Tr}_1^{n}\left(\frac{1}{w}\left(c+c^{-1}\right)\right)$, there are 
exactly $\frac{q}{2}$ elements $w$ of $\GF{q}^\star$ which satisfy \eqref{maincondition}
and then the number of the corresponding $t$'s is equal to $2\cdot
\frac{q}{2}=q.$

Now, let $z+z^{-1}\neq 0$ and set
\begin{equation}\label{s}
s=\frac{1}{z+z^{-1}}.
\end{equation} By Proposition~\ref{Aprop},
$t\in \mu_{q^2+1}$ if and only if $
\mathbf{Tr}_1^{2n}\left(T\right)=1$, i.e. letting
\begin{equation}\label{w}
w=s^{-1}\alpha,
\end{equation}
$$\mathbf{Tr}_1^{2n}\left(\frac{1}{s^{-1}+c\alpha s^{-1}}\right)=1.$$
It holds
\begin{align*}&\mathbf{Tr}_1^{2n}\left(\frac{1}{s^{-1}+c\alpha s^{-1}}\right)=\mathbf{Tr}_1^{n}\left(s\mathbf{Tr}_{n}^{2n}\left(\frac{1}{1+c\alpha}\right)\right)=\mathbf{Tr}_1^{n}\left(s\left(\frac{1}{1+c\alpha}+\frac{1}{(1+c\alpha)^q}\right)\right)
\\&=\mathbf{Tr}_1^{n}\left(s\left(\frac{1}{1+c\alpha}+\frac{1}{1+c^{-1}\alpha}\right)\right)=\mathbf{Tr}_1^{n}\left(s\left(\frac{\alpha(c+c^{-1})}{1+\alpha^2+\alpha(c+c^{-1})}\right)\right)
\\&=\mathbf{Tr}_1^{n}\left(\frac{s}{e(\alpha+\alpha^{-1})+1}\right)\end{align*}
where $e=\frac{1}{c+c^{-1}}$.  Note that $\mathbf{Tr}_1^{n}(e)=1$
and
 $\mathbf{Tr}_1^{n}(s)=0$ by
Proposition~\ref{Aprop} as $c\in \mu_{q+1}$ and $z\in \mu_{q-1}$.
Now we set \begin{equation}\label{alpha}
\frac{1}{\alpha+\alpha^{-1}}=\beta+e.
\end{equation}
Then,
$\mathbf{Tr}_1^{n}\left(\frac{s}{e(\alpha+\alpha^{-1})+1}\right)=\mathbf{Tr}_1^{n}\left(\frac{(\beta+e)s}{\beta}\right)=\mathbf{Tr}_1^{n}\left(s+\frac{es}{\beta}\right)=\mathbf{Tr}_1^{n}\left(\frac{es}{\beta}\right).$
Thus, the number of the pairs $(z,t)$ satisfying \eqref{Cond} is
\[
N=q+8\cdot \#\{(\beta,s)\in \GF{q}^2\mid
\mathbf{Tr}_1^{n}\left(\beta\right)=1,
\mathbf{Tr}_1^{n}\left(s\right)=0 \text{ and }
\mathbf{Tr}_1^{n}\left(\frac{es}{\beta}\right)=1 \}.
\]
Fix any $\beta\in \mathfrak{F}_1\setminus \{e\}$ and set
$\gamma=\frac{e}{\beta}$. Introduce the sets
$S_{0,i}:=\{s\in \GF{q}\mid \mathbf{Tr}_1^{n}\left(s\right)=0 \text{
  and } \mathbf{Tr}_1^{n}\left(\gamma s\right)=i\}$ for
$i\in \{0,1\}$. Observe that $S_{0,0}\cup S_{0,1}=\mathfrak{F}_0$ and
$S_{0,0}\neq \mathfrak{F}_0$ because every $s\in S_{0,0}$ satisfies
equation
$\gamma^{2^{n-1}}\mathbf{Tr}_1^{n}(s)+\mathbf{Tr}_1^{n}(\gamma s)=0$
which has degree $\frac{q}{4}=\frac{\#\mathfrak{F}_0}{2}$ in terms of
$s$ as $\gamma\neq 1$. So, $S_{0,1}\neq \emptyset$ and
$S_{0,1}=s_0+S_{0,0}$ for any element $s_0\in S_{0,1}$. Therefore
$\#S_{0,0}=\#S_{0,1}$. On the other hand
$\#S_{0,0}+\#S_{0,1}=\#\mathfrak{F}_0=\frac{q}{2}$. Thus,
$$ \#\{s\in \GF{q}\mid \mathbf{Tr}_1^{n}\left(s\right)=0 \text{ and
} \mathbf{Tr}_1^{n}\left(\gamma s\right)=1\}=\frac{q}{4}$$ and
$$N=q+8\cdot\frac{q}{4}\cdot\left(\frac{q}{2}-1\right)=q^2-q.$$ \qed
\end{proof}

\begin{remark}\label{rem-Scase} The proof of Lemma~\ref{Scase} gives more information than stated
in  Lemma~\ref{Scase} itself. In the case of $q^2-q$ solutions
in $\GF{q^4}$, the proof also gives an approach to finding these
solutions. Explicit solutions to  quadratic equations can be found in
\cite{KM19,MK20}.
\end{remark}

\subsection{Proof of Item~\ref{item:3}}\label{sec:proof-item-refitem:3}

In the below lemma, we prove the last item of
Theorem~\ref{thm:main-result}.

\begin{lemma}\label{Tcase}
We have
\[
\#\{x\in \GF{q^4}\mid
x^{q^3+q^2+q-1}+(x+1)^{q^3+q^2+q-1}=b\}=\begin{cases}2, \text{ if
$b\in \mathfrak{S}_2$,}\\
0, \text{otherwise.}
\end{cases}
\]
\end{lemma}
\begin{proof}
Let us assume that Equation~\eqref{ori_eq} has a solution $x\in
\GF{q^4}$. We can set
\begin{equation}\label{x}
x=\frac{1}{1+z\lambda t}
\end{equation} where $(z,\lambda,t)\in
\mu_{q-1}\times\mu_{q+1}\times \mu_{q^2+1}$. Since $b\in
\GF{q^4}\setminus \GF{q^2}$,  from the proof of Lemma~\ref{Scase} we
have $$\lambda\neq 1$$ and by Lemma~\ref{Fcase}
$$t\neq 1.$$ Then, Equality~\eqref{ori_eq}
can be rewritten as
$$\left(\frac{1}{1+z\lambda t}\right)^{q^3+q^2+q-1}\left(1+(z\lambda t)^{q^3+q^2+q-1}\right)=b.$$
Since $1+(z\lambda t)^{q^3+q^2+q-1}=(1+z\lambda^{-1}t^{-1})^2$ and
$(1+z\lambda t)^{q^3+q^2+q-1}=(1+ z\lambda^{-1}t^{-q})(1+z\lambda
t^{-1})(1+z\lambda^{-1} t^q)(1+z\lambda t)^{-1}$, this equality is
rewritten as
$$\frac{(1+z\lambda^{-1}t^{-1})^2(1+z\lambda t)}{(1+ z\lambda^{-1}t^{-q})(1+z\lambda t^{-1})(1+z\lambda^{-1}
t^q)}=b,$$ or
\begin{equation}\label{eq4}
\frac{(z+\lambda t)^2(z+\lambda^{-1} t^{-1})}{(z+\lambda t^{q})(
z+\lambda^{-1}t)(z+ \lambda t^{-q})}=b.
\end{equation}
Regarding
\begin{align*}
&z+\lambda^{-1} t^{-1}=(z+\lambda^{-1} t)^{q^2}, \\
&z+\lambda t^{q}=(z+\lambda^{-1} t)^{q},\\
&z+ \lambda t^{-q}=(z+ \lambda^{-1} t)^{q^3},
\end{align*}
Equality~\eqref{eq4} again can be rewritten as
\begin{equation}\label{eq5}
\frac{(z+\lambda t)^2}{( z+\lambda^{-1}t)^{q^3-q^2+q+1}}=b.
\end{equation}
Since  modulo $q^4-1$ it holds $(q^3-q^2+q+1)\cdot(q^2+1)\equiv 2
q\cdot(q^2+1)$, $(q^3-q^2+q+1)\cdot(q^2-1)\equiv 2\cdot(q^2-1)$ and
$(q^3-q^2+q+1)\cdot\frac{q^4-1}{q+1}\equiv
-2\cdot\frac{q^4-1}{q+1}$, by raising \eqref{eq5} to the
$(q^2+1)-$th, $(q^2-1)-$th, $\frac{q^4-1}{q+1}-$th powers we have
\begin{equation}\label{eq100}
\left(\frac{z+\lambda
t}{(z+\lambda^{-1}t)^q}\right)^{q^2+1}=\sqrt{b}^{q^2+1},
\end{equation}
\begin{equation}\label{eq8}
\left(\frac{z+\lambda
t}{z+\lambda^{-1}t}\right)^{q^2-1}=\sqrt{b}^{q^2-1},
\end{equation}
\begin{equation}\label{eq7}
\left((z+\lambda
t)(z+\lambda^{-1}t)\right)^{\frac{q^4-1}{q+1}}=\sqrt{b}^{\frac{q^4-1}{q+1}},
\end{equation}
respectively.
\end{proof}

Let $c=\frac{1}{\sqrt{b}}$ and $T=t+t^{-1}$. Then, $c^{q^2}+c\neq 0$
since $c\notin \GF{q^2}$ by the assumption, and
$$T\neq
T^q$$ since $\mathbf{Tr}_1^{2n}\left(\frac{1}{T}\right)=1$ via
Proposition~\ref{prop:decomposition}.

Now, from Equality~\eqref{eq8} it follows $\left(\frac{c(z+\lambda
t)}{z+\lambda^{-1}t}\right)^{q^2}=\frac{c(z+\lambda
t)}{z+\lambda^{-1}t}$, or $c^{q^2}(z+\lambda
t^{-1})(z+\lambda^{-1}t)=c(z+\lambda t)(z+\lambda^{-1}t^{-1}) $,
i.e.
\begin{equation}\label{eq9}
z^2+1=z\cdot \frac{c^{q^2}(\lambda t^{-1}+\lambda^{-1}t)+c(\lambda
t+\lambda^{-1}t^{-1})}{c^{q^2}+c}.
\end{equation}
And from Equality~\eqref{eq100}, it follows $c^{q^2+1}(z+\lambda
t)(z+\lambda t^{-1})=(z+\lambda t^{q})(z+\lambda t^{-q})$, i.e.
$(c^{q^2+1}+1)(z^2+\lambda^2)=\lambda z\cdot (c^{q^2+1}T+T^q).$ If
$c^{q^2+1}=1$, then this equality becomes $T=T^q$, a contradiction.
Therefore, $$c^{q^2+1}\neq1$$ and
\begin{equation}\label{eq10}
z^2+\lambda^2=\lambda z\cdot \frac{c^{q^2+1}T+T^q}{c^{q^2+1}+1}.
\end{equation}
Equalities~\eqref{eq9} and ~\eqref{eq10} give
\begin{equation}\label{eq11}
z=\frac{\lambda^2+1}{\lambda\cdot (A+B_1)+\lambda^{-1}\cdot B},
\end{equation}
where
\begin{align*}
&A=\frac{c^{q^2+1}T+T^q}{c^{q^2+1}+1}=T+\frac{T+T^q}{c^{q^2+1}+1},\\
&B_1=\frac{ct+c^{q^2}t^{-1}}{c+c^{q^2}}=t+\frac{c^{q^2}T}{c+c^{q^2}},\\
&B=\frac{ct^{-1}+c^{q^2}t}{c+c^{q^2}}=t^{-1}+\frac{c^{q^2}T}{c+c^{q^2}}=B_1+T.
\end{align*} Note that $A,B,B_1$ are in  $\GF{q^2}.$ Now, let us introduce some notation $$\alpha:=c^{q^2+1},
\beta:=c+c^{q^2}, \gamma:=\frac{c}{\beta},
\delta:=\left(\frac{\beta}{\alpha}\right)^{q-1}.$$ Note that
$$\alpha, \beta\in \GF{q^2}^*, \delta\in \mu_{q+1},
\gamma^{q^2}=1+\gamma.$$ Then,
$$A=\frac{\alpha
T+T^{q}}{\alpha+1}, B=\gamma t^{-1}+\gamma^{q^2}t, B_1=\gamma
t+\gamma^{q^2}t^{-1}.$$

From Equality \eqref{eq9} it follows
$$z+\frac{1}{z}=\lambda B_1+\lambda^{-1}B$$ and so
$\lambda(B_1+B^q)=\lambda^{-1}(B_1^q+B).$ If $B_1+B^q=0$ and
$B_1^q+B=0$, then $T+T^q=0$ which is a contradiction again since
$T\not\in\GF{q}$.  Thus, $$B_1+B^q\neq 0, B+B_1^q\neq 0$$ and we have
\begin{equation}\label{eq13}
\lambda=\left(\frac{B_1^q+B}{B_1+B^q}\right)^{\frac{1}{2}}.
\end{equation}

 At this step of our proof, it was shown that both $z$ and
$\lambda$ are uniquely determined given a value of $t$ by
\eqref{eq13} and \eqref{eq9}. Thus, to complete our proof, it remains only to prove that at most two $t$'s may exist.

 Let $M=z+z^{-1}$ and
$L=\lambda+\lambda^{-1}$. By using \eqref{eq13} and \eqref{eq9}, one
can get
$$L=\frac{B+B_1+B^q+B_1^q}{\sqrt{(B_1+B^q)(B+B_1^q)}}$$ and
  $$M=\lambda
B_1+\lambda^{-1}B=\frac{B^{q+1}+B_1^{q+1}}{\sqrt{(B_1+B^q)(B+B_1^q)}}.$$
Since Equality \eqref{eq7} is rewritten as
$$\left(c(z^2+t^2+ztL)\right)^{(q-1)(q^2+1)}=1$$ and here
$$(z^2+t^2+ztL)^{q^2+1}=(z^2+t^{-2}+zt^{-1}L)(z^2+t^2+ztL)=z^2(M^2+L^2+T^2+MLT),$$
it follows
$$\left(c^{q^2+1}(M^2+L^2+T^2+MLT)\right)^{q-1}=1.$$
By routine computations one can get
$M^2+L^2+T^2+MLT=M^2+L^2+(B+B_1)^2+ML(B+B_1)=\frac{(B+B_1+B^q+B_1^q)^2}{(B_1+B^q)(B+B_1^q)}\cdot
(1+BB_1).$ Since $T^{q^2}=T$ and $B^{q^2}=B$ (these can be directly
checked by using $t^{q^2}=t^{-1}$), it holds
$\frac{(B+B_1+B^q+B_1^q)^2}{(B_1+B^q)(B+B_1^q)}=\frac{(T+T^q)^2}{(B+B^q+T)(B+B^q+T^q)}\in
\GF{q}$ and therefore $\left(c^{q^2+1}(1+BB_1)\right)^{q-1}=1.$ By
the way,
\begin{equation}\label{eqBB}
1+BB_1=1+\frac{ct^{-1}+c^{q^2}t}{c+c^{q^2}}\cdot\frac{ct+c^{q^2}t^{-1}}{c+c^{q^2}}=\frac{c^{q^2+1}T^2}{(c+c^{q^2})^2}
\end{equation} and so we get $\left(\frac{c^{q^2+1}}{c+c^{q^2}}\cdot
T\right)^{q-1}=1$, i.e.
\begin{equation}\label{eqTT}
T^{q}=\delta T.
\end{equation}

Substituting \eqref{eq11} to \eqref{eq9} gives
$$\frac{\lambda^2+1}{\lambda\cdot (A+B_1)+\lambda^{-1}\cdot B}+\frac{\lambda\cdot (A+B_1)+\lambda^{-1}\cdot B}{\lambda^2+1}=\lambda B_1+\lambda^{-1}B,$$
or
$$\lambda^4+1+\lambda^2(A+B_1)^2+\lambda^{-2} B^2=(\lambda B_1+\lambda^{-1}B)(\lambda (A+B_1)+\lambda^{-1} B)(\lambda^2+1).$$
Since $(\lambda B_1+\lambda^{-1}B)(\lambda (A+B_1)+\lambda^{-1}
B)(\lambda^2+1)=(\lambda^2B_1(A+B_1)+\lambda^{-2}B^2+AB)(\lambda^2+1)=\lambda^4B_1(A+B_1)+\lambda^2(AB+B_1(A+B_1))+
B^2+AB+\lambda^{-2} B^2$, we get
$$\lambda^4\left(B_1^2+AB_1+1\right)+\lambda^2A(A+B+B_1)+B^2+AB+1=0,$$ or
$$\lambda^4\left(B^2+AB+1+T(A+T)\right)+\lambda^2A(A+T)+B^2+AB+1=0.$$ This
equality can be rewritten as follows
$$\lambda^2T(A+T)+A(A+T)+L^2\cdot(B^2+AB+1)=0$$ after dividing
the left side by $\lambda^2$. Then, by considering
$\lambda^2=\frac{B+B^q+\delta T}{B+B^q+T}$, $A=\frac{(\alpha+
\delta)T}{\alpha+1}$, $A+T=\frac{(\delta+1) T}{\alpha+1}\neq 0$ and
$L^2=\frac{(\delta^2+1)T^2}{(B+B^q+\delta T)(B+B^q+T)}$, it follows
$$\frac{B+B^q+\delta T}{B+B^q+T}+\frac{\alpha+\delta}{\alpha+1}+\frac{(\alpha+1)(\delta+1)(B^2+AB+1)}{(B+B^q+\delta
T)(B+B^q+T)}=0.$$ By rearranging this equality we get
$$(\alpha B+B^q+\alpha T)(\alpha B+B^q+\delta T)=(\alpha+1)^2,$$
i.e.
\begin{equation}\label{maineq}
(\alpha B_1+B^q)(\alpha B+B_1^q)=(\alpha+1)^2.
\end{equation}
Substituting
$1+BB_1\overset{\eqref{eqBB}}{=}\frac{\alpha T^2}{\beta^2}$,
$$B^{q+1}=\gamma^{q+1}
t^{-(q+1)}+\gamma^{q^2(q+1)}t^{q+1}+\gamma^{q^2+q}t^{1-q}+\gamma^{q^3+1}t^{q-1}$$
and $$B_1^{q+1}=\gamma^{q+1}
t^{q+1}+\gamma^{q^2(q+1)}t^{-(q+1)}+\gamma^{q^2+q}t^{q-1}+\gamma^{q^3+1}t^{1-q}$$
to this equality gives
$$
\left(\frac{\alpha^2}{\beta^2}+\frac{\alpha^{q-1}
\delta^2}{\beta^{2q}}\right)T^2+\left(\gamma^{q+1}+\gamma^{q^2(q+1)}\right)T_{q+1}+\left(\gamma^{q^2+q}+\gamma^{q^3+1}\right)T_{q-1}=0,
$$
where $T_{q-1}:=t^{q-1}+t^{1-q}$ and $T_{q+1}:=t^{q+1}+t^{-(q+1)}.$
Since $\frac{\alpha^2}{\beta^2}+\frac{\alpha^{q-1}
\delta^2}{\beta^{2q}}=\frac{\alpha^2}{\beta^2}+\frac{
1}{\alpha^{q-1}\beta^2}$ and $T_{q-1}=T_{q+1}^q$, it follows from this that
\begin{equation}\label{inittval}
\frac{\alpha^{q+1}+1}{\alpha^{q-1}\beta^2}T^2+(\gamma^{q+1}+\gamma^{q^2(q+1)})T_{q+1}+[(\gamma^{q+1}+\gamma^{q^2(q+1)})T_{q+1}]^q=0.
\end{equation}

Now, we express both $T_{q+1}$ and $T_{q+1}^q$ in terms of $t^2,
t^{-2}$ and $t^{2q}$. From
$(t+t^{-1})^q\overset{\eqref{eqTT}}{=}\delta (t+t^{-1})$, we have
\begin{equation}\label{Teq1}
t^{-(q+1)}+t^{q-1}=\delta(t^{-2}+1),
\end{equation}
and by raising this equality to the $q^3-$th power
\begin{equation}\label{Teq2}
t^{q+1}+t^{q-1}=\delta^q(t^{2q}+1).
\end{equation}
 Addition of Equalities \eqref{Teq1} and \eqref{Teq2}  yields
\begin{equation}\label{Teq3}
T_{q+1}=(\delta+\delta^q)+\delta t^{-2}+\delta^qt^{2q}.
\end{equation}
It holds
$T_{q+1}=(x^{q+1}+x^{-(q+1)})=(x^{q+1}+x^{-(q+1)})^{q^2}=T_{q+1}^{q^2}$
and so
\begin{equation}\label{Teq4}
T_{q+1}^q=T_{q+1}^{q^3}=(\delta+\delta^q)+\delta t^{2}+\delta^q
t^{2q}.
\end{equation}
Substitution of \eqref{Teq3} and \eqref{Teq4} to \eqref{inittval}
yields to
\begin{align*}
\delta^q\mathbf{Tr}_n^{4n}(\gamma^{q+1})t^{2q}+&
\left[(\gamma^{q+1}+\gamma^{q^2(q+1)})\delta+\frac{\alpha^{q+1}+1}{\alpha^{q-1}\beta^2}\right]
t^{-2}\\
&+\left[(\gamma^{q+1}+\gamma^{q^2(q+1)})^q\delta+\frac{\alpha^{q+1}+1}{\alpha^{q-1}\beta^2}\right]
t^{2}+(\delta+\delta^q)\mathbf{Tr}_n^{4n}(\gamma^{q+1})=0.
\end{align*}
Using $\gamma^{q^2}=\gamma+1$, $\delta^q=\delta^{-1}$,
$\gamma^{q+1}+\gamma^{q^2(q+1)}=\gamma+\gamma^q+1,$
$\left(\gamma^{q+1}+\gamma^{q^2(q+1)}\right)^q=\gamma+\gamma^q$ and
$\mathbf{Tr}_n^{4n}(\gamma^{q+1})=\gamma^{q+1}+\gamma^{q}(\gamma+1)+(\gamma+1)^{q+1}+\gamma(\gamma^q+1)=1$,
this is simplified as
\begin{align*}
t^{2q+2}+&
\delta^2\left[\gamma+\gamma^q+\frac{\alpha^{q+1}+1}{\delta\alpha^{q-1}\beta^2}\right]
t^{4}
\\
&+(1+\delta^2)t^2+\delta^2\left[1+\gamma+\gamma^q+\frac{\alpha^{q+1}+1}{\delta\alpha^{q-1}\beta^2}\right]=0.
\end{align*}
Letting
$U:=\gamma+\gamma^q+\frac{\alpha^{q+1}+1}{\delta\alpha^{q-1}\beta^2},$
that is
$$ t^{2q+2}+t^2+ \delta^2(U t^{4} +t^2+U+1)=0,
$$ or equivalently
\begin{equation}\label{teq1}t^2(t^2+1)^q=
\delta^2(t^2+1)(Ut^2+U+1),
\end{equation} i.e.
\begin{equation}\label{teq2}
t^2(t^2+1)^{q-1}= \delta^2(Ut^2+U+1).
\end{equation}
Raising \eqref{teq2} to the $q^2-$th power and after rearranging  we obtain
\begin{equation}\label{teq3}
t^{-2(q-1)}(t^2+1)^{q-1}= \delta^2((U+1)t^2+U).
\end{equation}
Dividing \eqref{teq2} by \eqref{teq3} yields
\begin{equation}\label{teq4}
t^{2q}= \frac{Ut^2+U+1}{(U+1)t^2+U}.
\end{equation}
By substituting \eqref{teq4} to \eqref{teq1} we have
$$
 t^2\cdot\frac{t^2+1}{(U+1)t^2+U}=\delta^2(t^2+1)(Ut^2+U+1),
$$
or
$$
t^4+\frac{1+\delta^{-2}}{U+U^2}t^2+1=0,
$$
i.e.
\begin{equation}\label{teqfinal}
T=t+\frac{1}{t}=\frac{1+\delta^q}{\sqrt{U+U^2}}.
\end{equation}
 After all, there are at most two possible $t$'s.
Moreover, easy computations give
$\frac{\alpha^{q+1}+1}{\delta\alpha^{q-1}\beta^2}=\frac{\alpha^{q+1}+1}{\beta^{q+1}},$
$U=\frac{(c^{q^2+q}+1)^{q^2+1}}{(c+c^{q^2})^{q+1}}$,
$U+U^2=\frac{\mathbf{N}_n^{4n}(c^{q+1}+1)}{(c+c^{q^2})^{2(q+1)}}\in
\GF{q}$ and
$1+\delta^q=\frac{\mathbf{Tr}_n^{4n}\left(c^{q^2+q+1}\right)}{c^{q^2+1}(c+c^{q^2})^q}$.
By Proposition~\ref{prop:decomposition}, Equation~\eqref{teqfinal}
has two solutions in $\mu_{q^2+1}$ if and only if
$\mathbf{Tr}_1^{2n}\left(\frac{\sqrt{U+U^2}}{1+\delta^q}\right)=1.$
Since
\begin{align*}
&\mathbf{Tr}_1^{2n}\left(\frac{\sqrt{U+U^2}}{1+\delta^q}\right)=\mathbf{Tr}_1^{2n}\left(\frac{c^{q^2+1}}{c+c^{q^2}}\cdot
\frac{\sqrt{\mathbf{N}_n^{4n}(c^{q+1}+1)}}{\mathbf{Tr}_n^{4n}\left(c^{q^2+q+1}\right)}\right)\\
&=\mathbf{Tr}_1^{n}\left(\mathbf{Tr}_n^{2n}\left(\frac{c^{q^2+1}}{c+c^{q^2}}
\right)\cdot\frac{\sqrt{\mathbf{N}_n^{4n}(c^{q+1}+1)}}{\mathbf{Tr}_n^{4n}\left(c^{q^2+q+1}\right)}\right)\\
&=\mathbf{Tr}_1^{n}\left(\frac{\mathbf{Tr}_n^{4n}\left(c^{q^2+q+1}\right)}{(c+c^{q^2})^{q+1}}
\cdot\frac{\sqrt{\mathbf{N}_n^{4n}(c^{q+1}+1)}}{\mathbf{Tr}_n^{4n}\left(c^{q^2+q+1}\right)}\right)\\
&=\mathbf{Tr}_1^{n}\left(\mathbf{N}_n^{4n}\left(\frac{c^{q+1}+1}{c+c^{q^2}}\right)\right)=\mathbf{Tr}_1^{n}\left(\mathbf{N}_n^{4n}\left(\frac{b^{q+1}+1}{b+b^{q^2}}\right)\right),
\end{align*}
it follows that Equation~\eqref{teqfinal} has two solutions in
$\mu_{q^2+1}$ if and only if
\begin{equation}\label{condition}
\mathbf{Tr}_1^{n}\left(\mathbf{N}_n^{4n}\left(\frac{b^{q+1}+1}{b+b^{q^2}}\right)\right)=1.
\end{equation}

It is obvious that $\lambda$ given by \eqref{eq13} lies in
$\mu_{q+1}$ as  $B,B_1\in\GF{q^2}.$  Furthermore, it is easy to check
$(1+\delta^q)^{q-1}=\delta$ and that $T$ given by \eqref{teqfinal}
satisfies \eqref{eqTT}. Now, we will show that $z$ given by
\eqref{eq11}, \eqref{teqfinal} and \eqref{condition} lies actually
in $\GF{q}$, i.e.
$$\frac{\lambda^2+1}{\lambda\cdot (A+B_1)+\lambda^{-1}\cdot B}=\frac{\lambda^{-2}+1}{\lambda^{-1}\cdot (A^q+B_1^q)+\lambda\cdot B^q}.$$
This equality holds if $A+A^q+B_1+B_1^q=\lambda^2B^q+\lambda^{-2}B.$
By substituting \eqref{eq13} to this equality and rearranging in
regard that
$A+A^q=(T+T^q)(1+\frac{1}{\alpha+1}+\frac{1}{(\alpha+1})^q)$ and
$B+B^q+B_1+B_1^q=T+T^q$, one can get
\begin{equation}\label{verieq}
\left(\frac{1}{\alpha+1}+\frac{1}{(\alpha+1})^q\right)(B_1+B^q)(B+B_1^q)=BB_1+(BB_1)^q.
\end{equation}
One can also confirm that \eqref{maineq} is satisfied if
 \eqref{teqfinal} and \eqref{condition} are true by plotting the above discussion back. Therefore, by \eqref{maineq} one has
 $(B_1+B^q)(B+B_1^q)=(\alpha+1)\left(BB_1+\frac{\alpha+1+BB_1^q}{\alpha}\right)$.
Then, direct computation using \eqref{eqBB} verifies \eqref{verieq}.

Thus, it follows that a triple $(z,\lambda, t)$ given
\eqref{teqfinal}, \eqref{condition}, \eqref{eq13} and \eqref{eq11}
lies indeed in $\mu_{q-1}\times\mu_{q+1}\times \mu_{q^2+1}$ and
satisfies \eqref{eq9}, \eqref{eq10}, so also \eqref{eq100},
\eqref{eq8} and hence \eqref{eq5}.
\qed

\begin{remark}\label{rem:Tcase}
  Moreover, for $b\in \mathfrak{S}_2$, the two solutions in $\GF{q^4}$
  to $x^{q^3+q^2+q-1}+(x+1)^{q^3+q^2+q-1}=b$ are given by \eqref{x},
  \eqref{teqfinal}, \eqref{eq13} and \eqref{eq11}.
\end{remark}

From Lemma~\ref{Fcase}, Lemma~\ref{Scase} and Lemma~\ref{Tcase}, it
follows, which may be of independent interest,
\begin{corollary}
$\#\left\{b\in \GF{q^4}\setminus \GF{q^2} \mid
  \mathbf{Tr}_1^{n}\left(\mathbf{N}_n^{4n}\left(\frac{b^{q+1}+1}{b+b^{q^2}}\right)\right)=1\right\}=\frac{q^3(q-1)}{2}.$
 \end{corollary}



\section{Conclusion}\label{Sec-Conclusion}
In this paper, we have been interested in solving the equation
$X^{2^{3n}+2^{2n}+2^{n}-1}+(X+1)^{2^{3n}+2^{2n}+2^{n}-1}=b$ in
$\GF{2^{4n}}$. To this end, we have presented a powerful approach that gives the number of solutions of the equation and explicitly determines the set of $b$'s for which the equation has $i$ solutions for any positive integer $i$ simultaneously.

As an immediate  consequence of our results, we prove that the
problem stated in \cite{BCCDK20} as a conjecture is valid leading to a complete determination of the differential spectrum of the related power function $x^d$  defined over $\GF{2^{4n}}$ where $d=2^{3n}+2^{2n}+2^{n}-1$. Our approach differs from the nice one adopted by Li et al., which focuses on solving the conjecture and determining the related differential spectrum only. We believe that our algebraic technique  could be helpful to solve other similar interesting problems.  In particular, solving $X^{2^{(k-1)n}+2^{(k-2)n}+\cdots +2^{n}-1}+(X+1)^{2^{(k-1)n}+2^{(k-2)n}+\cdots +2^{n}-1}=b$ in
$\GF{2^{kn}}$ for other values of $k$ would  be an interesting future work.



\section*{Acknowledgement} The first author thanks Dok Nam Lee for 
useful discussions  on this work.

\end{document}